\documentclass[twocolumn, showpacs, preprintnumbers, pre]{revtex4}
\usepackage{amssymb, amsmath, amssymb, amsfonts}
\usepackage{graphicx}

\newcommand{\be}{\begin{equation}}
\newcommand{\ee}{\end{equation}}
\newcommand{\BE}{\begin{eqnarray}}
\newcommand{\EE}{\end{eqnarray}}
\newcommand{\BM}{\begin{multline}}
\newcommand{\EM}{\end{multline}}

\begin{document}

\title{The linear noise approximation for reaction-diffusion systems on networks}
\author{Malbor Asllani$^{1}$, Tommaso Biancalani$^{2}$, Duccio Fanelli$^{3}$, Alan J. McKane$^{2}$} 
\affiliation{$^{1}$Dipartimento di  Scienza e Alta Tecnologia, Universit\`{a} degli Studi dell'Insubria, via Valleggio 11, 22100 Como, Italy} 
\affiliation{$^{2}$Theoretical Physics, School of Physics and Astronomy,
University of Manchester, Manchester M13 9PL, U.K.}
\affiliation{$^{3}$Dipartimento di Fisica e Astronomia, Universit\`{a} di Firenze and INFN, via
Sansone 1 50019 Sesto Fiorentino, Florence, Italy }

\begin{abstract}
Stochastic reaction-diffusion models can be analytically studied on complex networks using the linear noise approximation. This is illustrated through the use of a specific stochastic model, which displays traveling waves in its deterministic limit. The role of stochastic fluctuations is investigated and shown to drive the emergence of stochastic waves beyond the region of the instability predicted from the deterministic theory. Simulations are performed to test the theoretical results and are analyzed via a generalized Fourier transform algorithm. This transform is defined using the eigenvectors of the discrete Laplacian defined on the network. A peak in the numerical power spectrum of the fluctuations is observed in quantitative agreement with the theoretical predictions.
\end{abstract}

\maketitle

\section{Introduction}
\label{sec:Intro}
Pattern formation in reaction-diffusion models is a subject which finds applications in many distinct fields, including ecology ~\cite{mimura,maron,baurmann,riet}, biology~\cite{meinhardt, harris, maini, bhat, miura} and chemistry~\cite{prigogine,castets,quyang}. In his seminal paper, Turing~\cite{turing} pointed out that a small perturbation of a homogeneous state in a reaction-diffusion system can, due to an instability, rapidly grow to eventually yield stable non-homogeneous patterns. The vast majority of studies devoted to investigating the emergence of Turing patterns use deterministic partial differential equations to model the reactions and diffusion of the constituents, which are characterized by continuous spatial distributions.

These ideas have recently been generalized in two distinct directions. First, the concept of a Turing instability can be extended to embrace systems defined on complex networks~\cite{nakao}. This is an important step forward~\cite{vespignani}, which should eventually shed novel light onto the mechanisms that drive self-organization on networks~\cite{colizza, satorras, colizzavespignani}. Second, Turing patterns have also been observed and analyzed in models with a finite number of constituents. In this case the reaction-diffusion processes are described by individual-based models which take into account the intrinsic discreteness of the system. Stochastic effects are therefore present, and ultimately stem from the finite size of the population of elementary constituents. In this paper we build on earlier work~\cite{ourwork}, to bring these two features together, and study stochastic patterns on networks. Specifically, we predict the existence of stochastic waves on networks, and show how they may be described analytically. We illustrate this on a specific model, but it should be clear from the analysis that we present that the ideas and techniques are generally applicable.

At first sight it might appear surprising that stochastic effects are important in reaction-diffusion systems, which after all consist of a large number of constituents. However, the fluctuations due to the discreteness of these constituents can amplify through resonant effects and so yield macroscopically ordered patterns, both in time~\cite{mckanePRL, dipatti} and in space~\cite{goldenfeld, biancalani, woolley}. Stochastic Turing patterns \cite{biancalani} (also termed quasi-Turing patterns \cite{goldenfeld}) can appear in a region of the parameter
space for which the homogeneous fixed point is predicted to be stable from a deterministic linear stability analysis. Similarly, stochastic waves~\cite{tommasoOnde} have been observed in reaction-diffusion models defined on a regular lattice. Importantly, the effect of fluctuations arising from this discreteness can not only be seen in numerical simulations, but also analytically understood by expanding the governing master equation within the so-called Linear Noise Approximation (LNA) scheme.

Starting from this setting, in \cite{ourwork} stochastic Turing patterns have been predicted, and numerically observed, on a scale free network, so extending the conclusions of \cite{nakao}, beyond the deterministic setting. Here we report analytical progress by carrying out the LNA for systems in which the population is distributed over a set of nodes, which are connected to each other in some way. Each node hosts a large number of individuals, so this is a metapopulation model~\cite{hanski} --- a collection of populations allowing some exchange between them. Examples are: in ecology, where individuals reside on patches and may migrate to other patches that are nearby~\cite{hanski}; in island models of evolutionary theory, where individuals carrying certain alleles may migrate to other islands~\cite{maruyama}; in epidemiology, where the nodes are cities connected by commuters who carry disease~\cite{ganna} and in reaction kinetics, where the nodes are compartments in which chemical reactions take place~\cite{joe}.

More concretely, we shall consider a stochastic version of the Zhabotinsky model~\cite{zhab}, introduced into a static scale-free network. This latter is created via the preferential attachment probability rule~\cite{barabasi}. The power spectrum of fluctuations is analytically calculated by developing and systematizing the LNA technique to network-based applications. A localized peak for the power spectrum signals the presence of stochastic travelling waves, a prediction that we confirm with stochastic simulations~\cite{gillespie}. The power spectrum is calculated from a generalized Fourier transform, the standard plane waves found in a spatial context being replaced by the eigenvectors of the discrete Laplacian operator defined on the network. To benchmark theory and simulations, we have therefore implemented and tested a numerical routine which handles the generalized Fourier analysis. This is a diagnostic tool that could prove useful beyond the specific case study, by guiding the unbiased search for structured patterns on a network topology~\cite{kouvaris}.  

The paper is organized as follows. In the next section we will introduce the stochastic model and discuss the basic steps of the LNA analysis on a graph. The technical aspects of the presentation will be relegated to specific Appendices. In Section III the linear stability analysis for the model in its deterministic limit is carried out. Numerical simulations are performed to show that the deterministic wave manifests itself as a localized peak in the power spectrum, as obtained from the generalized Fourier transform. In Section IV we derive the power spectrum of fluctuations which points to the existence of traveling waves, seeded by inherent stochasticity, outside the region of deterministic instability. Stochastic simulations confirm the validity of the theory. In the final Section we sum up and conclude.

\section{Model definition and the linear noise approximation (LNA)}
\label{sec:Model}

The reaction scheme that we will investigate was introduced by Zhabotinsky et al. to study travelling waves arising from destabilization of a homogeneous state \cite{zhab}. The scheme involves molecules of three chemical species: $X$, $Y$ and $Z$ --- that we will also respectively call the first, second and third species. The molecules are placed on the nodes of a network composed of $\Omega$ nodes, each of which has a finite volume $V$. We label a molecule of species $X$ located on the $i$-th node with $X_i$; $Y_i$ and $Z_i$ are similarly defined. The number of molecules of type $X_i$, $Y_i$ and $Z_i$ are denoted by $x_i$, $y_i$ and $z_i$, respectively. The $\Omega$-dimensional vectors: $\mathbf{x}=\left(x_1,...,x_{\Omega}\right)$, $\mathbf{y}=\left(y_1,...,y_{\Omega}\right)$ and $\mathbf{z}=\left(z_1,...,z_{\Omega}\right)$, specify the state of the system. Within each node, the molecules interact through the following reaction scheme:
\begin{eqnarray}\label{eq:react}
X_i+2Y_i &\xrightarrow{\;\;\;c_1\;\;\;}& 2Y_i,\nonumber\\
X_i+2Y_i &\xrightarrow{\;\;\;c_2\;\;\;}& X_i+3Y_i,\nonumber\\
2Z_i &\xrightarrow{\;\;\;c_3\;\;\;}& X_i+2Z_i,\nonumber\\
Y_i &\xrightarrow{\;\;\;c_4\;\;\;}& \varnothing,\\
X_i &\xrightarrow{\;\;\;c_5\;\;\;}& X_i+Z_i,\nonumber\\
Z_i &\xrightarrow{\;\;\;c_6\;\;\;}& \varnothing,\nonumber\\
X_i &\xrightarrow{\;\;\;c_7\;\;\;}& \varnothing,\nonumber\\
\varnothing &\xrightarrow{\;\;\;c_8\;\;\;}& Y_i.\nonumber
\end{eqnarray}

The reaction rates are denoted by $c_1, c_2,...,c_8$. As explained in \cite{zhab} they are all constant except $c_7$ that is given by $c_7 = c'_7/\left(g+\frac{x_i}{V}\right)$, with $g = 10^{-4}$.

The structure of the network is described by the $\Omega \times \Omega$ adjacency matrix, $W$. This is a symmetric matrix whose elements, $W_{ij}$, is equal to one if node $i$ is connected to node $j$, and zero otherwise. The molecules can migrate between two connected nodes as specified by the diffusion reactions:
\be \label{eq:diff}
X_i \xrightarrow{\;\;\;d_1\;\;\;} X_j,\quad Y_i\xrightarrow{\;\;\;d_2\;\;\;} Y_j, \quad Z_i\xrightarrow{\;\;\;d_3\;\;\;} Z_j.
\ee
The constants $d_1$, $d_2$ and $d_3$ are the diffusion coefficients. 

The construction of a stochastic model proceeds by assigning a transition rate $\mathrm T(\mathbf{x}', \mathbf{y}', \mathbf{z}' | \mathbf{x}, \mathbf{y}, \mathbf{z})$ to each reaction. They indicate the probability per unit of time to transit from state $(\mathbf{x}, \mathbf{y}, \mathbf{z})$ to state $(\mathbf{x}', \mathbf{y}', \mathbf{z}')$. To lighten the notation, we only write the components of the vectors which refer to molecules that take part in a reaction in the transition rates. Invoking mass action, the transition rates associated with reactions \eqref{eq:react} read \cite{vanKampen}:
\begin{eqnarray} \label{eq:Treact}
\mathrm T_1(x_i-1|x_i ) &=& c_1 \frac{x_i}{V}\frac{y_i^2}{V^2}, \nonumber \\
\mathrm T_2(y_i+1|y_i) &=& c_2 \frac{x_i}{V}\frac{y_i^2}{V^2}, \nonumber \\
\mathrm T_3(x_i+1|x_i) &=& c_3 \frac{z_i^2}{V^2}, \nonumber \\
\mathrm T_4(y_i-1|y_i) &=& c_4 \frac{y_i}{V},   \\
\mathrm T_5(z_i+1|z_i) &=& c_5 \frac{x_i}{V}, \nonumber \\
\mathrm T_6(z_i-1|z_i) &=& c_6 \frac{z_i}{V}, \nonumber \\
\mathrm T_7(x_i-1|x_i) &=& c_7 \frac{x_i}{V}, \nonumber  \\ 
\mathrm T_8(y_i+1|y_i) &=& c_8. \nonumber 
\end{eqnarray}
In a similar way the transition rates for the diffusion reactions \eqref{eq:diff} are given by:
\BE \label{eq:Tdiff}
\mathrm T_9(x_i-1,x_j+1|x_i,x_j) &=& d_1 \frac{x_i}{V}, \nonumber  \\
\mathrm T_{10}(y_i-1,y_j+1|y_i,y_j) &=& d_2 \frac{y_i}{V},  \\
\mathrm T_{11}(z_i-1,z_j+1|z_i,z_j) &=& d_3\frac{z_i}{V}. \nonumber 
\EE
As the dynamics is Markovian, the probability density function (PDF) that the system is in state $(\mathbf{x},\mathbf{y},\mathbf{z})$ at time $t$,  $\mathrm P(\mathbf{x},\mathbf{y},\mathbf{z},t)$, satisfies the master equation: 
\begin{multline} \label{eq:master}
\!\!\!\!\!\!\frac{\partial}{\partial t}\mathrm P(\mathbf{x},\mathbf{y},\mathbf{z},t)=\!\!\!\!\!\!\!\!\!\!\!\!\sum_{(\mathbf{x}',\mathbf{y}',\mathbf{z}'\neq\mathbf{x},\mathbf{y},\mathbf{z})} \!\!\!\!\!\![\mathrm T(\mathbf{x},\mathbf{y},\mathbf{z}|\mathbf{x}',\mathbf{y}',\mathbf{z}')\mathrm P(\mathbf{x}',\mathbf{y}',\mathbf{z}',t)\\-\mathrm  T(\mathbf{x}',\mathbf{y}',\mathbf{z}'|\mathbf{x},\mathbf{y},\mathbf{z})\mathrm P(\mathbf{x},\mathbf{y},\mathbf{z},t)].
\end{multline}
This is the fundamental equation that governs the dynamics of the system. 

The LNA can be applied by carrying out the van Kampen expansion for the master equation \cite{vanKampen}. This begins with changing variables from $(x_i, y_i, z_i)$ to $(\xi_{1,i}, \xi_{2,i}, \xi_{3,i})$, where $i=1,\ldots,\Omega$:
\BE
\frac{x_i}{V} =\phi_i+\frac{\xi_{1,i}}{\sqrt{V}},\quad \frac{y_i}{V}=\psi_i+\frac{\xi_{2,i}}{\sqrt{V}},\quad \frac{z_i}{V}&=&\eta_i+\frac{\xi_{3,i}}{\sqrt{V}}. 
\nonumber \\
\label{eq:ansatz}
\EE
The functions $\phi_i(t)$, $\psi_i(t)$ and $\eta_i(t)$ describe the concentrations of each chemical species in the deterministic limit, that is, obtained by letting $V \rightarrow \infty$. In this limit the system is not subject to fluctuations. As shown in Appendix \ref{sec:A}, the deterministic concentrations satisfy the following system of ordinary differential equations:
\begin{eqnarray}\label{eq:det}
\dot \phi_i&=& - c_1\phi_i\psi_i^2 + c_3 \eta_i^2- c'_7 \frac{\phi_i}{g + \phi_i} + d_1\sum_{j=1}^{\Omega} \Delta_{ij}\phi_j, \nonumber\\
\dot \psi_i&=& c_2 \phi_i\psi_i^2 - c_4 \psi_i+ c_8 + d_2\sum_{j=1}^{\Omega} \Delta_{ij}\psi_j, \\
\dot \eta_i&=& c_5 \phi_i - c_6 \eta_i + d_3\sum_{j=1}^{\Omega} \Delta_{ij}\eta_j. \nonumber
\end{eqnarray}
Hereafter, a dot above a symbol indicates the time derivative taken with respect to the rescaled time, $\tau = t/V$.  The symbol $\Delta_{ij}$ denotes the Laplacian operator defined on the network and reads: 

\be \label{eq:lap}
\Delta_{ij}=W_{ij}-k_i\delta_{ij},
\ee
where $k_i$ is the connectivity of node $i$, $k_i = \sum_{j=1}^{\Omega} W_{ij}$. This form of the Laplacian operator \cite{vulpiani} reflects our specific choice for the microscopic reaction rates \eqref{eq:Tdiff}. Other choices are possible which would yield modified Laplacians \cite{sneppen}. Working in the context of the proposed formulation, $\Delta_{ij}$ is symmetric, a feature of which we shall take advantage of 
when performing the generalized Fourier analysis described below.

For finite volume $V$, the system is subject to intrinsic noise; these fluctuations perturb the solutions of the deterministic model \eqref{eq:det}, which describes the dynamics of the model in the limit $V \to \infty$. Within the LNA, the fluctuations are Gaussian and given by a linear Fokker-Planck equation. Before turning to discuss the role played by stochastic fluctuations, we will start by focusing on the deterministic scenario. We will in particular derive the conditions under which self-organized patterns of the wave type emerge. The next section is devoted to this issue.   

\section{Pattern formation in the deterministic limit} \label{sec:det}
The analysis of pattern formation for system \eqref{eq:det} defined on a regular lattice in the continuum limit has been already carried out in \cite{zhab}. Here, we review some of the results of \cite{zhab}, before moving on to discuss how the network affects the pattern formation. Throughout our analysis we have used a scale-free network generated with the Barab\'{a}si-Albert preferential attachment algorithm \cite{barabasi}, with $\Omega$ nodes and mean degree $\langle k \rangle$. 

We first establish contact with the notation adopted in \cite{zhab} by making the following choices: $c_1=c_3=m$, $c'_7=a\,m$, $c_2=c_4=n$, $c_8=b\, n$ and $c_5=c_6=1$. As in \cite{zhab}, we fix some of the parameters: $a=0.9$, $b=0.2$ and $d_1=d_2=0$. We also set $d_3=0.8$. The parameters $(m,n)$ can be freely adjusted and select different dynamical regimes.

The system of differential equations \eqref{eq:det} admits three fixed points \cite{zhab}. One of these corresponds to the extinction of both $X$ and $Z$ species and is always stable. Another one is a saddle. The last one is non-trivial, and its stability depends on the values of $(m,n)$. It is around this point in the two-dimensional plane defined by $m$ and $n$ that the pattern formation is investigated. The concentrations $(\phi^*, \psi^*, \eta^*)$  at the fixed point are independent of ($m,n$) and can be numerically determined. 

Patterns arise when $(\phi^*, \psi^*, \eta^*)$ becomes unstable with respect to inhomogeneous perturbations \cite{cross}. To look for instabilities, we introduce small deviations from the fixed point, $(\delta \phi_i, \delta \psi_i, \delta \eta_i)$, and linearise system \eqref{eq:det} around it:
\BE \label{eq:linstab}
	\begin{pmatrix} \delta \dot \phi_i \\ \delta \dot \psi_i \\  \delta \dot \eta_i \end{pmatrix} = \sum_{j=1}^{\Omega} \left( \mathcal {M^*}^{(NS)} \delta_{ij} +   \mathcal {M^*}^{(SP)} \Delta_{ij} \right) \cdot \begin{pmatrix} \delta \phi_j \\ \delta \psi_j \\  \delta \eta_j \end{pmatrix}.
\EE
The explicit expressions for the matrices $\mathcal {M^*}^{(NS)}$ and $\mathcal {M^*}^{(SP)}$ are given in Appendix \ref{sec:A} (the label NS stands for ``non-spatial'' and SP for ``spatial''). For a regular lattice, the Fourier transform is usually employed to solve the above linear equations. This analysis needs to be adapted in the case of a system defined on a network. To this end we follow the approach of  \cite{nakao, ourwork} and start by defining the eigenvalues and eigenvectors of the matrix $\Delta$:
\be \label{eq:lapegv}
\sum_{j=1}^{\Omega} \Delta_{ij} v^{(\alpha)}_j =  \Lambda^{(\alpha)} v^{(\alpha)}_i, \quad \alpha = 1,\ldots,\Omega. 
\ee  
Since the Laplacian is symmetric, the eigenvalues $\Lambda^{(\alpha)}$ are real and the eigenvectors $v^{(\alpha)}$ form an orthonormal basis. It can actually be proven that for a case of a Barab\'{a}si-Albert network the $\Lambda^{(\alpha)}$ are negative and non-degenerate \cite{nakao, ourwork}. We can now define a transform based on the eigenvectors $v^{(\alpha)}$ which takes the role that the Fourier transform took on for a regular lattice. This leads to the following transforms which will be used throughout the remainder of the paper:
\BE \label{eq:trans}
		f_j(\tau) &=& \frac{1}{2\pi} \int_{-\infty}^{\infty}d\omega \sum_{\alpha=1}^{\Omega} \tilde f_\alpha(\omega) v^{(\alpha)}_j e^{- \mathrm i \omega \tau}, \nonumber \\
		\tilde f_\alpha(\omega) &=& \int_{0}^{\infty} d \tau \sum_{j=1}^{\Omega} f_j(\tau) v^{(\alpha)}_j e^{\mathrm i \omega \tau},
\EE  
where $f_j(\tau)$ is any function of the nodes and of time. This is a standard Fourier transform in time, but with the spatial Fourier modes replaced by the eigenvectors of the network Laplacian. If the network is a regular lattice, the transform \eqref{eq:trans} reduces to a standard Fourier transform for discrete space. From now on the index $\alpha$ is used to label the variable conjugate to the nodes. Using this definition, one can define the power spectrum as $P(\omega, \Lambda^{(\alpha)})=|\tilde f_\alpha(\omega)|^2$. 

Applying the transform \eqref{eq:trans} to Eq.~\eqref{eq:linstab} yields the following linear equation:
\be
	-\mathrm i\, \omega \begin{pmatrix} \delta \phi_\alpha \\ \delta \psi_\alpha \\  \delta \eta_\alpha \end{pmatrix} = \left( \mathcal {M^*}^{(NS)} + \mathcal {M^*}^{(SP)} \Lambda^{(\alpha)} \right) \cdot \begin{pmatrix} \delta \phi_\alpha \\ \delta \psi_\alpha \\  \delta \eta_\alpha \end{pmatrix},
\ee
that is now decoupled in the nodes and in time and thus readily solvable. The matrix $\mathcal {M^*}^{(NS)} + \mathcal {M^*}^{(SP)} \Lambda^{(\alpha)}$ for a given $\alpha$, is a $3\times 3$ matrix whose eigenvalues characterize the response of system \eqref{eq:det} to external perturbations. The eigenvalue with the largest real part will be denoted by $\lambda^{\rm max}(\Lambda^{(\alpha)})$. If $\text{Re}[\lambda^{\rm max}(\Lambda^{(\alpha)})] > 0$ the fixed point is unstable and the system exhibits a pattern whose spatial properties are encoded by $\Lambda^{(\alpha)}$. This is the analog of the wavelength for a spatial pattern in a a system defined on a regular lattice; it is customarily written $\Lambda^{(\alpha)} \equiv -k^2$ in this case. When the imaginary part of the eigenvalue,  $\text{Im}[\lambda^{\rm max}(\Lambda^{(\alpha)})]$, is different from zero, the pattern oscillates in time \cite{cross}. A system unstable for $\Lambda^{(\alpha)} \ne 0$ and $\text{Im}[\lambda^{\rm max}(\Lambda^{(\alpha)})] \ne 0$ is said to undergo a wave-instability and the emerging patterns consist of travelling waves.


\begin{figure*}[t]
\includegraphics[scale=0.30]{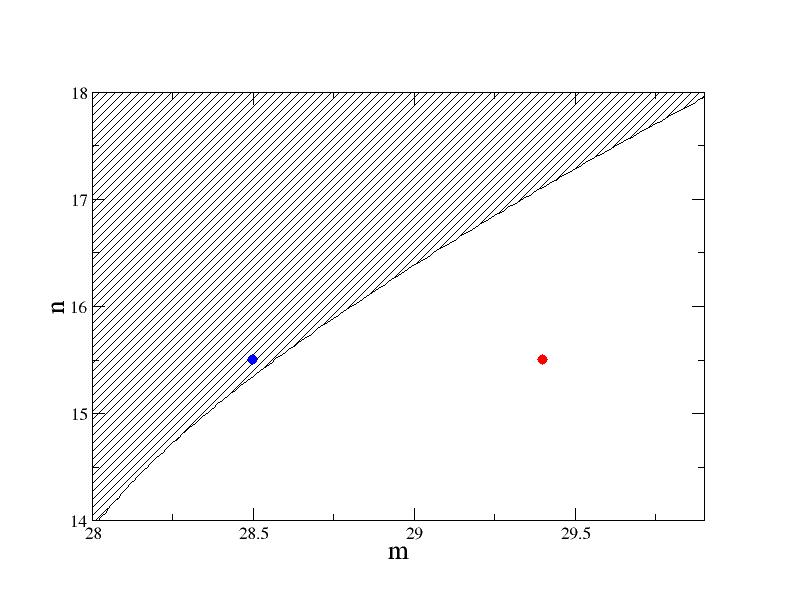}\begin{large}(\textbf{a})\end{large}\hspace*{0.5cm}
\includegraphics[scale=0.30]{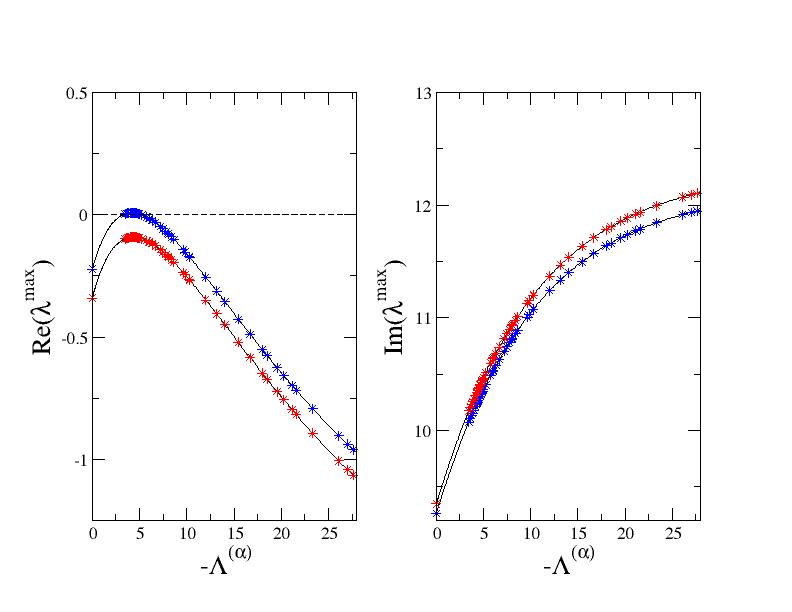}\begin{large}(\textbf{b})\end{large}
\caption{(Colour online) ($\textbf{a}$) The shaded region (left) delineates the wave instability domain in the $(m,n)$ plane for the Zhabotinsky model with $a = 0.9$, $b=0.2$, $d_1=d_2=0$ and $d_3 = 0.8$. The blue (online) circle falls inside of the region of wave instability and is at the point $(28.5,15.5)$. The red (online) circle is outside the ordered region and is at the point $(29.4,15.5)$. ($\textbf{b}$) Real and imaginary parts (right) of $\lambda^{\rm max}$ are plotted as a function of both the discrete modes $-\Lambda^{(\alpha)}$ of the network Laplacian (symbols) and their spatial analogs $-k^2$ (solid line). The parameters used are $(m,n) = (28.5,15.5)$, (blue online, upper curves) and $(m,n) = (29.4,15.5)$, (red online, lower curves) and correspond respectively to the two points identified in panel ($\textbf{a}$). The scale-free network employed in this analysis has $50$ nodes with a mean degree $\langle k \rangle=10$. The fixed point of the system is found to be  
$(\phi^* \approx 1.1308, \psi^* \approx 0.5787, \eta^* \approx 1.1308)$.
\label{fig:disp}}
\end{figure*}


In Fig.~\ref{fig:disp}, left panel, the domain of instability is shown as a shaded region in the plane $(m,n)$. The fixed point $(\phi^*, \psi^*, \eta^*)$ is stable for fixed $n$ when $m > m_c$. At $m=m_c$ a wave instability sets it and travelling waves are found to occur for $m < m_c$. The real and imaginary parts of the eigenvalues $\lambda^{\rm max}$ are depicted in the right panel, as a function of $-\Lambda^{(\alpha)}$, for two choices of the parameters $(m,n)$, for which the system is respectively stable and unstable. The circles in the left panel of Fig.~\ref{fig:disp} indicate these two choices.

Since the system is defined on a network, the emerging patterns present two main differences as compared to those obtained for the case of conventional reaction-diffusion models defined on the continuum. First, only some of the wavelengths, $\Lambda^{(\alpha)}$, are allowed. This is due to the fact that the solutions of Eq.~\eqref{eq:lapegv} form a discrete set; such a feature also occurs for systems defined on periodic lattices. In this latter case however, the wavelengths are equally spaced and proportional to the lattice spacing. By contrast, for systems defined on a complex network, there is no clear periodic structure and the wavelengths are clustered or irregularly distributed, as displayed in panel ($\textbf{b}$) of Fig.~\ref{fig:disp}. The second unusual trait has to do with the shape of the patterns. In a reaction-diffusion system defined on a regular lattice, each point of of the lattice has a concentration which assumes a value significantly different from that characterising the homogeneous fixed point. However, for a network, only a fraction of nodes have concentrations which are significantly different from that of the homogeneous fixed point. The fraction that are differentiated in this way depends on the connectivity of the nodes and on the ratio of the diffusion coefficients \cite{nakao}. This feature cannot be simply understood from linear stability analysis, as it relates to the localization of the Laplacian eigenvectors in large networks, a property that has been recently investigated in this context in~\cite{menzinger}.

\begin{figure}[h]
\begin{center}
\includegraphics[scale=0.21]{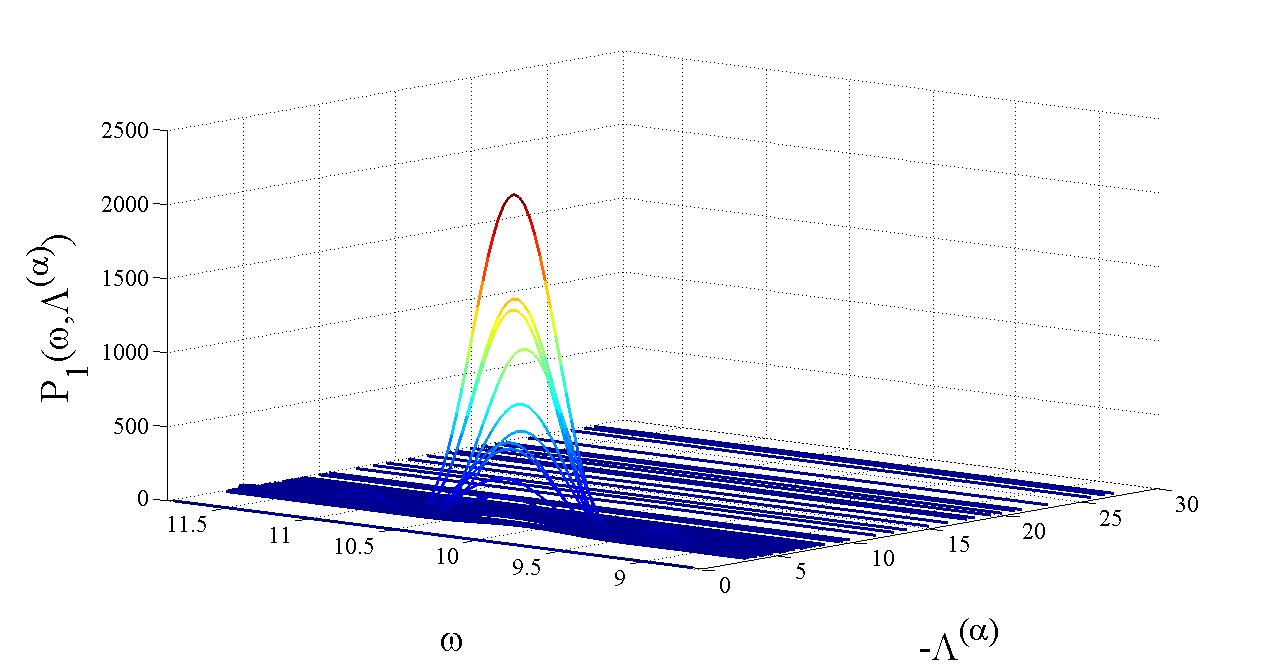} \\\vspace*{0.5cm}
\includegraphics[scale=0.21]{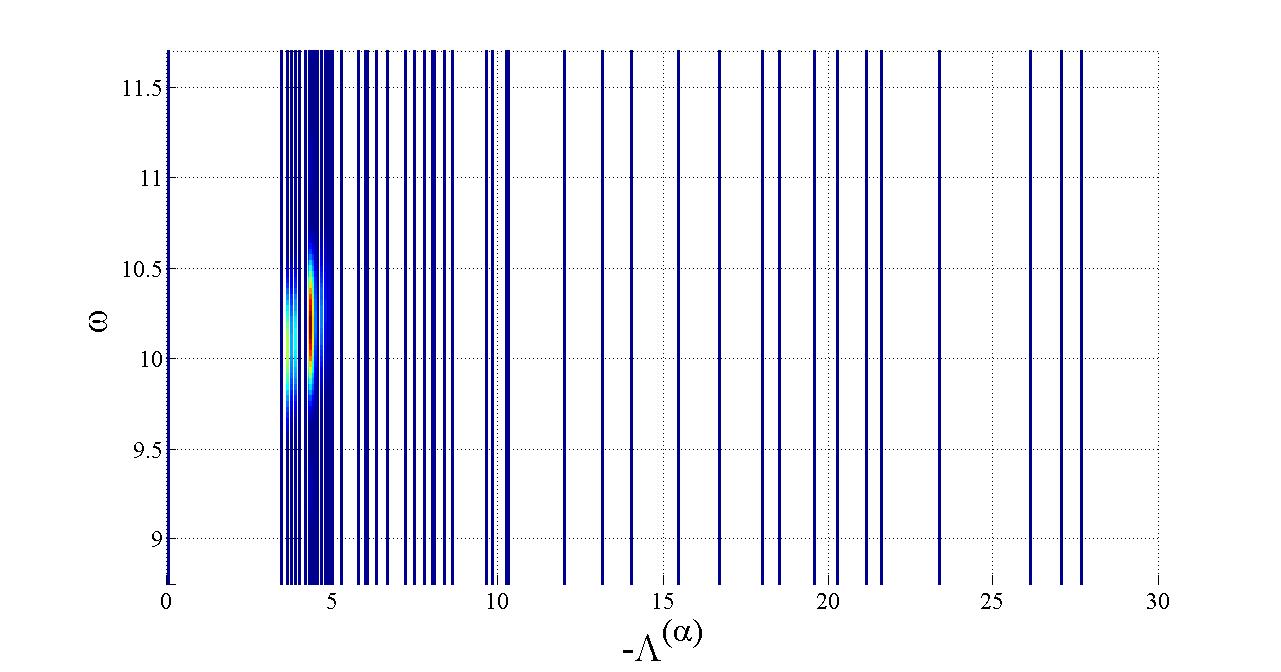} 
\end{center}
\caption{(Color online) Upper panel: power spectrum of the concentration $\phi_i(t)$, for a choice of the parameters that correponds to the unstable configuration of figure \ref{fig:disp}($\textbf{a}$) (blue circle online). The power spectrum is constructed from the generalized Fourier transform (\ref{eq:trans}), using as an input the numerical solution of the deterministic equations (\ref{eq:det}). A peak is seen for ($\omega, \Lambda^{(\alpha)}$)  $\simeq$ ($10,-5$), confirming the validity of the linear stability analysis and revealing the presence of a traveling wave in the time series. Lower panel: a two-dimensional projection of the power spectrum is displayed. Recall that the power spectrum is defined over a discrete, non-uniform support in $\Lambda^{(\alpha)}$.
\label{fig:PS_det}}
\end{figure}

In Fig.~\ref{fig:PS_det} the power spectrum of the concentration $\phi_i(\tau)$ is plotted for a choice of the parameters that correspond to the leftmost circle (blue online) in Fig.~\ref{fig:disp}($\textbf{a}$). A peak is displayed for ($\omega, \Lambda^{(\alpha)}$)  $\simeq$ ($10,-5$), in complete agreement with the predictions of the linear stability analysis. Similar results are obtained for the other concentrations $\psi_i(\tau)$ and $\eta_i(\tau)$. Thus, the generalized Fourier algorithm based on Eqs.(\ref{eq:trans})  can be effectively employed to resolve complex patterns that develop on networks. This is a valuable tool which, we believe, could prove useful for the many applications where the dynamics on a network is well known to be central, from neuroscience to epidemics.


The next section is entirely dedicated to the study of stochastic patterns, aiming at generalizing the deterministic picture of Fig.~\ref{fig:disp}. By applying the LNA, we will demonstrate that stochastic waves exist in a region of the parameter space for which the deterministic analysis predicts a stable homogeneous fixed point. The presence of stochastically driven patterns will be revealed by an analytical calculation of the power spectra of fluctuations. The theoretical predictions will then be validated by reconstructing the power spectrum from the stochastic time series. Properties of the patterns that in the deterministic picture depend on the eigenvectors, such as localization, will not be addressed in the present work.

\section{Power spectra of fluctuations and stochastic patterns} \label{sec:PS}

While in the deterministic limit a study of the eigenvalues reveals the range of parameter values for which patterns are expected to occur, this prediction is not conclusive for systems that are subject to noise. Simulations of the master equation have shown that patterns arise even for parameter values for which the underlying fixed point is stable, provided that the system is sufficiently close to an instability. The corresponding patterns have been called stochastic patterns \cite{biancalani} or quasi-patterns \cite{goldenfeld}. The LNA allows one to gain analytical insight into the mechanism that yields such patterns \cite{lugo, deanna, biancalani, goldenfeld}. The method has been extended in \cite{ourwork} to the case of a reaction-diffusion system on a network. Here we shall develop the method further and provide the first evidence for the spontaneous emergence of {\it stochastic waves} (or {\it quasi-waves}) on a network. 

The stochastic perturbations about the fixed point are described by the variables $\xi_{r,i}$ ($r=1,2,3$ and $i=1,\ldots,\Omega$) as defined in \eqref{eq:ansatz} with $\phi_i = \phi^*$, $\psi_i = \psi^*$ and $\eta_i = \eta^*$. 

\begin{figure}[h]
\begin{center}
\includegraphics[scale=0.21]{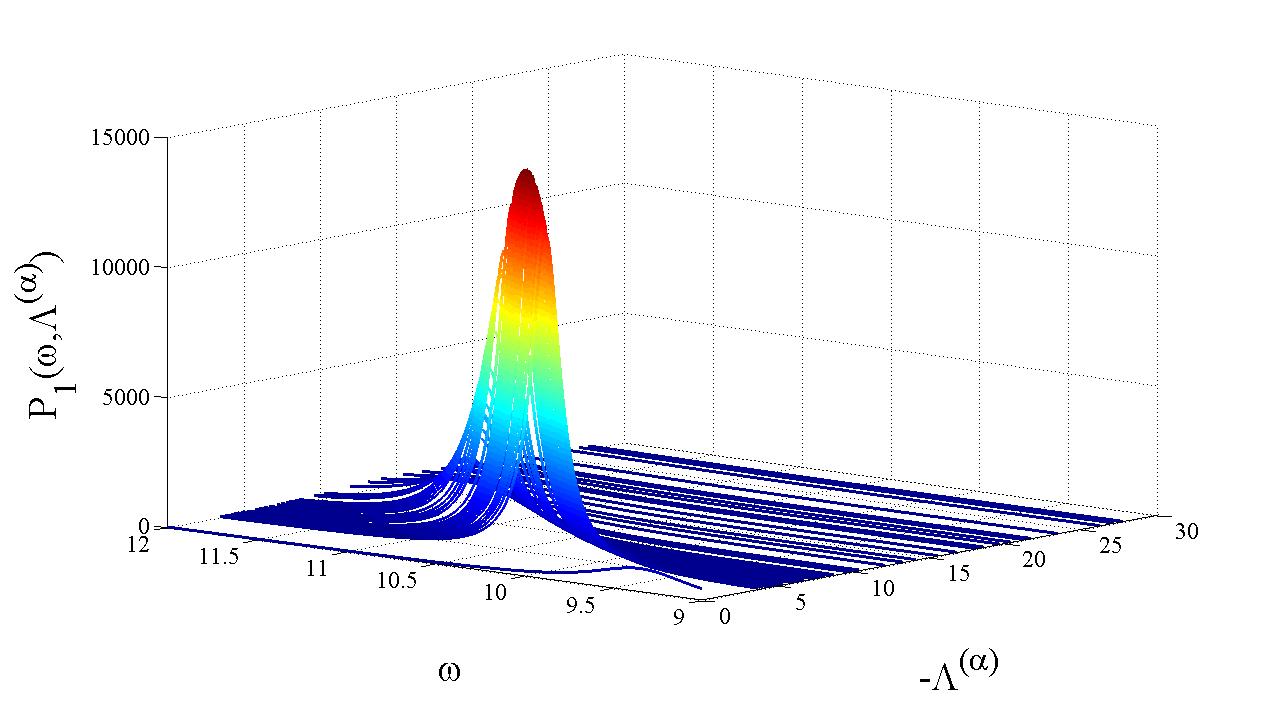}\\\hspace*{0.5cm}
\includegraphics[scale=0.21]{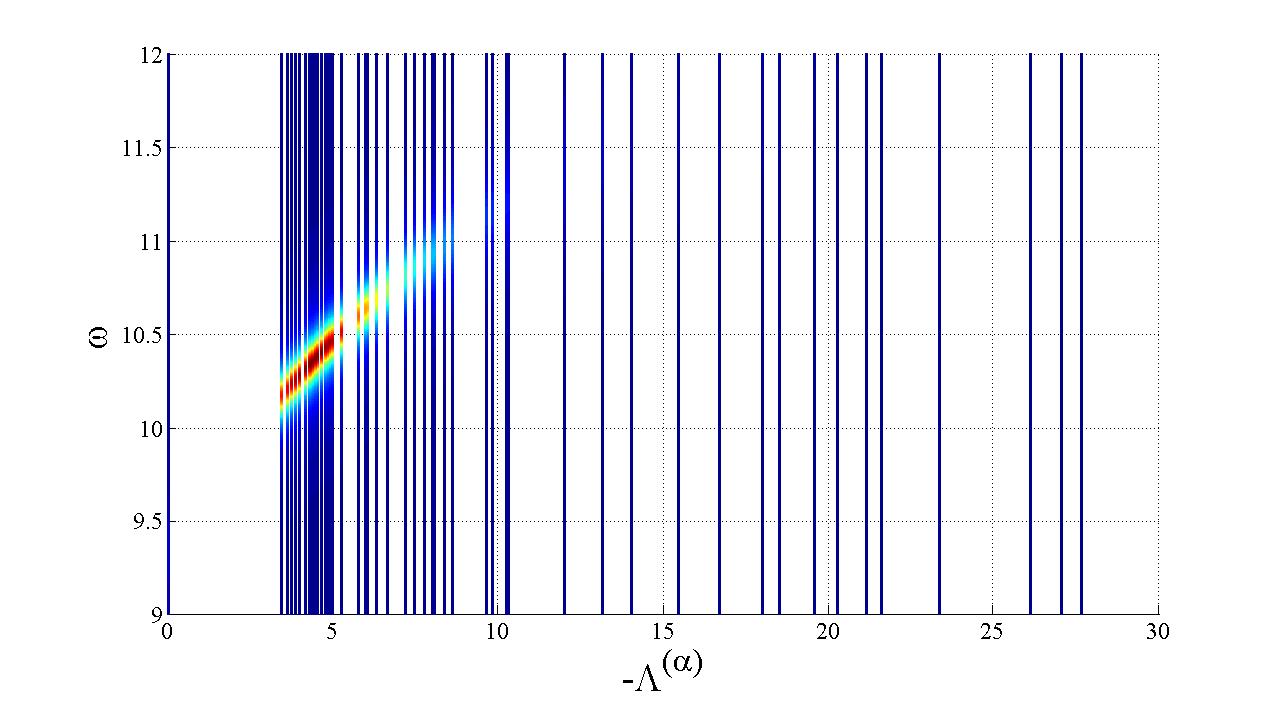}
\caption{(Color online) Upper panel: analytical power spectrum of the fluctuations, plotted as a function of the continuum frequency $\omega$  and the discrete wavelength $\Lambda_\alpha$. The parameters ($m,n$) are chosen so as to fall outside the region of deterministic order, i.e.~as indicated by the rightmost circle (red online) of Fig. \ref{fig:disp}($\textbf{a}$). The other parameters are set to the values specified in the caption of Fig.~\ref{fig:disp}. Lower panel: a two-dimensional projection of the power spectrum is displayed.
\label{fig:PS_out_an}}
\end{center}
\end{figure}

As discussed in Appendix~\ref{sec:A}, the role of fluctuations can be quantified by use of the van Kampen system-size expansion, which is equivalent to assuming the LNA. The quantity $1/\sqrt{V}$ acts as the small parameter in the perturbative expansion: at the leading order, the macroscopic deterministic equations (\ref{eq:det}) are recovered. At the next-to-leading order one obtains the Fokker-Planck equation (\ref{eq:fp2}) for the distribution of stochastic fluctuations $\Pi(\boldsymbol \xi_1,\boldsymbol \xi_2,\boldsymbol \xi_3, t)$. To quantify the impact of the stochastic components of the dynamics, the power spectrum of fluctuations is utilized. It is defined as: 
\be \label{eq:ps}
P_r(\omega, \Lambda^{(\alpha)}) = \langle \tilde \xi_{r,\alpha}^c(\omega) \tilde \xi_{r,\alpha}(\omega) \rangle = \langle | \tilde \xi_{r,\alpha}(\omega)|^2 \rangle,
\ee 
where $\tilde \xi^c_{r,\alpha}(\omega)$ is the complex conjugate of $\tilde \xi_{r,\alpha}(\omega)$, which is given by transforming $\xi_{r,i}(\tau)$ via the generalized transform \eqref{eq:trans}. The average $\langle \cdot \rangle$ is performed over many realizations of the stochastic dynamics. Following the strategy outlined in Appendix \ref{sec:B}, one can then derive an analytic expression for the power spectrum of fluctuations. This is equation (\ref{eq:analps}). Once the parameters of the model have been assigned, it is therefore possible to calculate the power spectrum of fluctuations and look for signatures of emerging self-organized structures. In Fig.~\ref{fig:PS_out_an} the analytical power spectrum for species $X$ is plotted for a choice of parameters that corresponds to the rightmost circle in Fig.~\ref{fig:disp} ($\textbf{a}$), namely outside the region for which the deterministic waves occur. 

    
As can be seen, the power spectrum of fluctuations is characterized by a localized peak for $(\omega_M, \Lambda^{(\alpha)}_M)$. Therefore, species $r=1$ oscillates with an angular frequency $\omega_M$ and, at the same time, displays a pattern at wavelength $\Lambda^{(\alpha)}_M$. Stochastic waves, or quasi waves, are hence predicted to occur, in a region of the parameter plane for which the homogeneous fixed point is stable, according to the deterministic picture.  In other words, stochastic corrections, stemming from finite size, and, as such, endogenous to the system under scrutiny, can eventually produce macroscopically ordered structures. 

\begin{figure}[t]
\begin{center}
\includegraphics[scale=0.19]{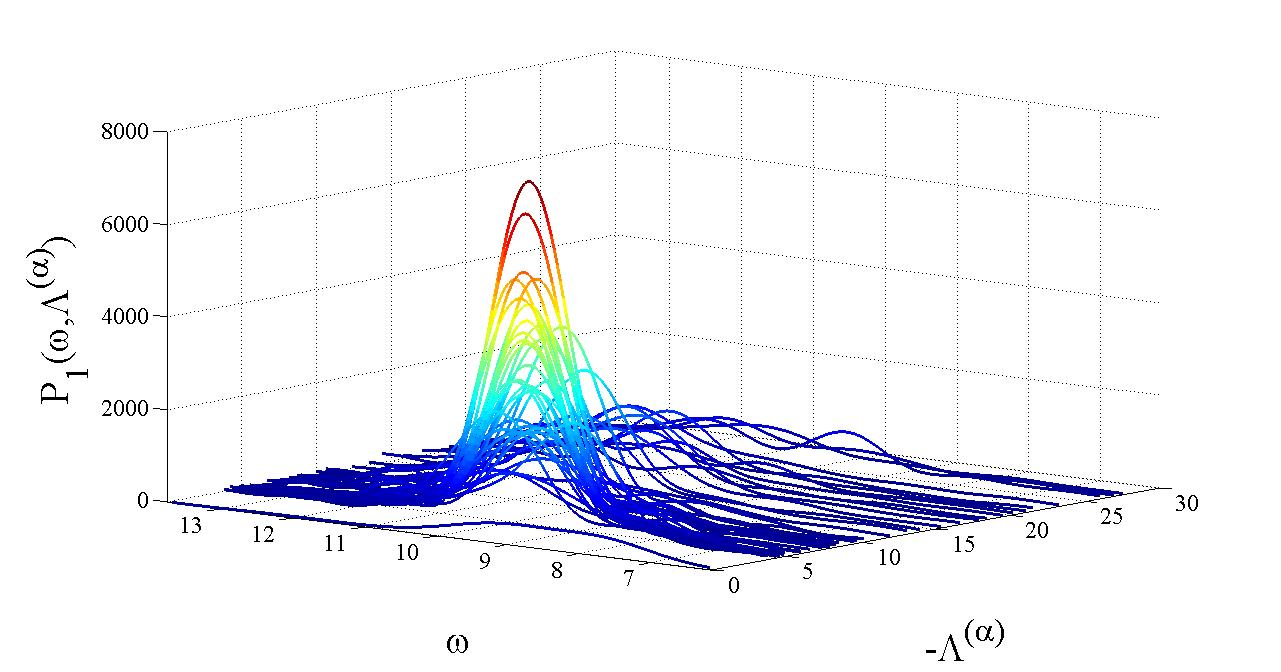}\\\hspace*{0.5cm}
\includegraphics[scale=0.19]{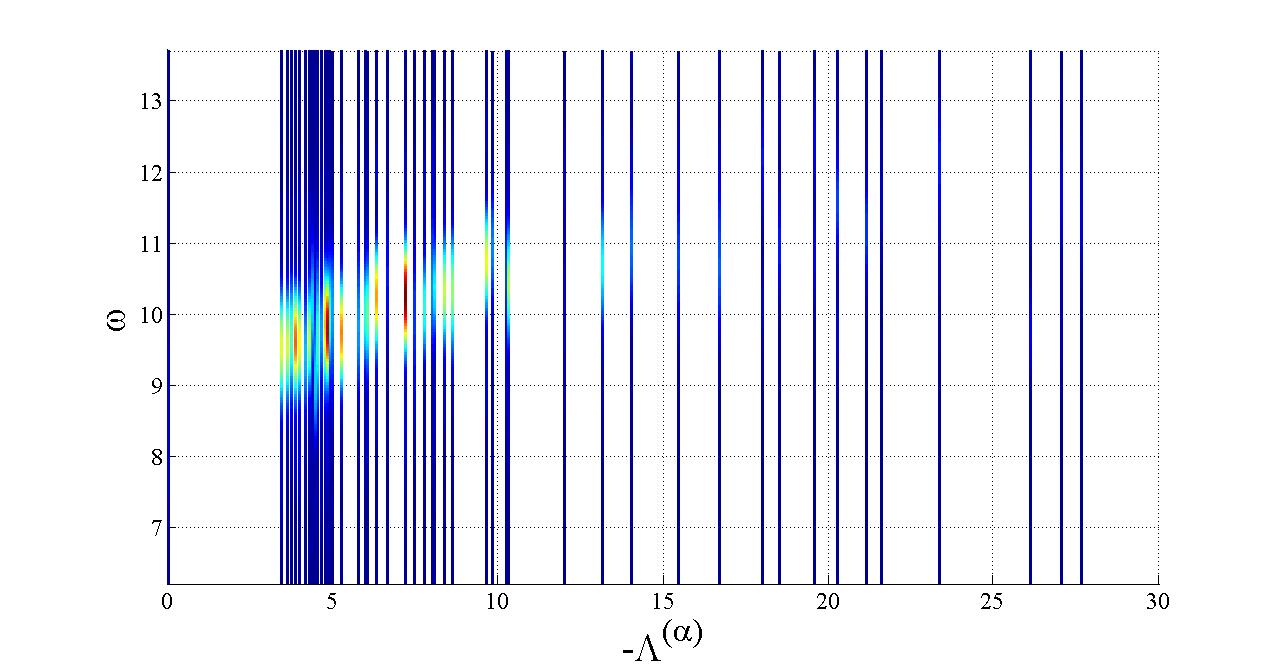}
\caption{(Color online) Upper panel: numerical power spectrum of the fluctuations obtained by simulating the stochastic dynamics via the Gillespie algorithm. The power spectum is calculated by using the generalized Fourier transform~\eqref{eq:trans} and by averaging over $40$ independent realizations. The parameters are the same as in Fig.~\ref{fig:PS_out_an}. Here $V=10^4$. Lower panel: two-dimensional projection of the power spectrum.
\label{fig:PS_out_num}}
\end{center}
\end{figure}

To test the correctness of the theoretical prediction we carried out stochastic simulations of the processes (\ref{eq:react}) and (\ref{eq:diff}) using the Gillespie algorithm~\cite{gillespie}. The numerical power spectrum is reconstructed by applying the generalized transform \eqref{eq:trans} to the time series, and averaging over independent realizations of the stochastic dynamics. The result is shown in Fig.~\ref{fig:PS_out_num} and is seen to agree with the theoretically-predicted spectrum.  The location of the maximum is captured by the theory, as well as the characteristic shape of the profile. 


\section{Conclusions}

Pattern formation has been extensively studied in the literature and with reference to a wide variety of problems. Typically, the concentrations of the species involved are assumed to obey partial differential equations. The conditions under which an instability occurs follow from standard linear stability analysis around a stable homogeneous fixed point. Recently, the emergence of steady state inhomogeneous patterns has been also studied for deterministic reaction-diffusion models defined on a network, generalizing the concept of a Turing instability to this important new area.  

Deterministic models represent, however, an idealized approach to the phenomenon being investigated: they omit stochastic fluctuations that need to be included when dealing with finite populations of interacting elements. Finite size corrections result in intrinsic stochastic perturbations which undergo amplification, and through a resonant mechanism eventually yield self-organized patterns. 

In this paper, we have further developed the theory of stochastic patterns for reaction-diffusion systems defined on a network. The analysis is based on a systematic and general application of the linear noise approximation scheme. Stochastic traveling waves are predicted to exist, and numerically observed in a region of the parameter plane for which the deterministic set of partial differential equations converges to a stable homogeneous solution. The analysis is carried out for a specific system, the stochastic analog of the deterministic model introduced in~\cite{zhab}. The techniques discussed are however general, and can be readily adapted to any reaction diffusion model defined on a network. To benchmark theory and simulations we have also developed, and successfully tested, a numerical algorithm that performs the generalized Fourier transform, employed in the analytical derivation. This transform decomposes the signal along the eigenvectors of the discrete Laplacian operator, tailoring the analysis to the network under consideration, so allowing the spectral properties of the emerging patterns to be fully characterized.  

\begin{acknowledgments}
T. B. thanks the EPSRC (UK) for partial financial support. D.F. acknowledges financial support from Ente Cassa di Risparmio di Firenze and the Program Prin2009 funded by MIUR (Italy). 
\end{acknowledgments}

\bibliographystyle{apsrev4-1}
\bibliography{bibliography}

\appendix
\section{Van Kampen system-size expansion} \label{sec:A}
In the main text, the van Kampen system-size expansion was used to approximate the master equation \eqref{eq:master} by a deterministic system of ordinary differential equations --- that describes the macroscopic evolution of the concentrations --- together with a linear Fokker--Planck equation which characterizes the fluctuations around the macroscopic solution. This Appendix details the calculations using the system-size expansion.

In the following (and throughout the paper) the indexes $i$ and $j$ refer to the nodes of the network and range from $1$ to $\Omega$. The indexes $r$ and $s$ label the chemical species and range from one to three. Finally, the $\Omega$-dimensional vectors, such as $\mathbf x$, $\mathbf y$ and $\mathbf z$, are displayed in bold.

Master equations, such as \eqref{eq:master}, can be rewritten by making use of step operators, $\epsilon^{\pm}_{r,i}$, which represent the creation/destruction of a molecule of species $r$ at node $i$. For instance, for species $X$, they act on a general function $f \left(\mathbf{x}, \mathbf{y}, \mathbf{z}\right)$ by
\begin{equation} \label{eq:step}
\epsilon_{1,i}^{\pm} f \left(...,x_i,..., \mathbf{y}, \mathbf{z}\right)=f\left(..., x_i\pm 1, ..., \mathbf{y}, \mathbf{z}\right).
\end{equation}
The master equation \eqref{eq:master} then reads:
\begin{widetext} 
\BE \label{eq:master2}
  \frac{\partial}{\partial t} \mathrm  P(\mathbf{x},\mathbf{y},\mathbf{z},t) = \sum_{i=1}^{\Omega}\Biggl[ &&(\epsilon_{1,i}^+-1)\mathrm T_1(x_i-1|x_i) + (\epsilon_{2,i}^--1) \mathrm  T_2(y_i+1|y_i) + 
 (\epsilon_{1,i}^--1)\mathrm  T_3(x_i+1|x_i) +  
 (\epsilon_{2,i}^+-1) \mathrm T_4(y_i-1|y_i) + \nonumber \\
 &&(\epsilon_{3,i}^--1)\mathrm T_5(z_i+1|z_i) +  
 (\epsilon_{3,i}^+-1)\mathrm  T_6(z_i-1|z_i) + 
 (\epsilon_{1,i}^+-1) \mathrm T_7(x_i-1|x_i) + 
 (\epsilon_{2,i}^--1)\mathrm  T_8(y_i+1|y_i) + \nonumber \\
&& \sum_{j=1}^{\Omega}W_{ij}\big[(\epsilon_{1,i}^+\epsilon_{1,j}^--1) \mathrm  T_{9}(x_i-1,x_j+1|x_i,x_j) 
+ (\epsilon_{2,i}^+\epsilon_{2,j}^--1) \mathrm  T_{10}(y_i-1,y_j+1|y_i,y_j) + \\
&& (\epsilon_{3,i}^+\epsilon_{3,j}^--1) \mathrm  T_{11}(z_i-1,z_j+1|z_i,z_j)\big] \Biggr] \mathrm{P}(\mathbf{x},\mathbf{y},\mathbf{z},t) \nonumber.
\EE
\end{widetext}

We now apply the change of variable \eqref{eq:ansatz}. In the new variables, the step operators admit an expansion in powers of $V^{-1}$ \cite{vanKampen}. The LNA corresponds to the truncation of the expansion at second order, namely:
\be \label{eq:stepexp}
\epsilon_{r,i}^{\pm} \approx 1 \pm \frac{1}{\sqrt{V}}\partial_{\xi_{r,i}}+\frac{1}{2V}\partial^2_{\xi_{r,i}}.
\ee
The left-hand side of Eq.~\eqref{eq:master2} can be expressed in terms of the PDF of the new variables, $\Pi(\boldsymbol \xi_1,\boldsymbol \xi_2,\boldsymbol \xi_3, t) = \mathrm P\left(\mathbf x \left( \boldsymbol \phi(t), \boldsymbol \xi_1
\right), \mathbf y \left( \boldsymbol \psi(t), \boldsymbol \xi_2 \right) , \mathbf z\left( \boldsymbol \eta(t), \boldsymbol \xi_3 \right), t \right)$. This implies that
\be \label{eq:melhs}
	\frac{\partial \mathrm P}{\partial t} = \partial_t \Pi -\sqrt V \left( \nabla_{\boldsymbol \xi_1} \Pi \cdot \partial_t \boldsymbol \phi + \nabla_{\boldsymbol \xi_2} \Pi \cdot \partial_t \boldsymbol \psi + \nabla_{\boldsymbol \xi_3} \Pi \cdot \partial_t \boldsymbol \eta \right).
\ee
In the above equation there are terms which are either $O(1)$ or $O(\sqrt V)$. By contrast, the right-hand side of Eq.~\eqref{eq:master2} contains $O(V^{-1/2})$ and $O(V^{-1})$ terms. They can be balanced by rescaling time by $\tau = t/V$. Collecting together the terms of the same order and setting their sum at each order to zero gives, at the leading order, the deterministic system \eqref{eq:det}. Likewise, the next-to-leading order yields the Fokker-Planck equation for the fluctuations:
\BE \label{eq:fp2}
\partial_\tau \Pi &=& \sum_{i=1}^{\Omega}  \left( - \sum_{r=1}^3 \partial_{\xi_{r,i}}  \left( \mathcal M_{r,i}\Pi \right) \right. \nonumber \\
&& \left. + \frac{1}{2}\sum_{r,s=1}^3 \sum_{j=1}^{\Omega}  
\partial_{\xi_{s,i}} \partial_{\xi_{r,j}} 
\left( \mathcal B_{rs,ij}\Pi \right) \right).
\EE

Equation \eqref{eq:fp2} is linear as the matrices $\mathcal M$ and $\mathcal B$ do not depend on $\boldsymbol \xi_r$, with $r=1,2,3$. However, they do depend on the trajectory $\boldsymbol \phi(\tau)$, $\boldsymbol \psi(\tau)$, $\boldsymbol \eta(\tau)$ that should be chosen beforehand among the solutions of \eqref{eq:det}. 

The form of the matrices $\mathcal M$ and $\mathcal B$ follow from the expansion of the transition rates \eqref{eq:Treact} and \eqref{eq:Tdiff}. For illustrative purposes, here we shall discuss only the first of the transition rates \eqref{eq:Tdiff}, $\mathrm T_9$, explicitly. The contribution to matrix $\mathcal B$ associated with this term, labeled $\mathcal B^{(9)}$, is found to be
\be
	\mathcal B^{(9)}_{rs,ij} = d_1 \delta_{rs,11} \left( 2 k_i \phi_i - W_{ij} \left( \phi_i + \phi_j \right) \right).
\ee
Clearly, the only non-zero entry is for $r=s=1$, since the rate $\mathrm T_9$ involves only the $X$ species. The other diffusion rates, $\mathrm T_{10}$ and $\mathrm T_{11}$, yield similar contributions for respectively $r=s=2$ and $r=s=3$, with diffusion coefficients and concentrations corresponding to the diffusing species. The contributions arising from the transition rates for the reactions \eqref{eq:Treact} follow in a similar fashion. 

In most applications, the main point of interest is to study the fluctuations around a fixed point. This is certainly so our case, as we aim to characterize the pattern that originates from a small perturbation of the fixed point $(\phi^*, \psi^*, \eta^*)$. We therefore substitute $\phi_i(\tau) = \phi^*$, $\psi_i(\tau) = \psi^*$ and $\eta_i(\tau) = \eta^*$ and label by $\mathcal M^*$ and $\mathcal B^*$ the matrices evaluated at the fixed point.

From the form of the reaction rates it is clear that the following decompositions hold~\cite{ourwork}: 
\BE \label{eq:decom}
\mathcal{M^*}_{sr, ij} &=& \mathcal{M^*}_{sr}^{(NS)}\delta_{ij} + \mathcal{M^*}_{sr}^{(SP)}\Delta_{ij}, \nonumber \\ 
\mathcal{B^*}_{sr, ij} &=& \mathcal{B^*}_{sr}^{(NS)}\delta_{ij} + \mathcal{B^*}_{sr}^{(SP)}\Delta_{ij}. 
\EE
The non-spatial part (NS) refers to the transition rates \eqref{eq:Treact}, whereas the spatial contribution (SP) refers to the transition rates \eqref{eq:Tdiff}. 

We end by giving the elements of the matrix $\mathcal M^*$ and $\mathcal B^*$.
The elements of $\mathcal M^*$ are
\BE \label{eq:Mfp}
\mathcal{M^*}_{11}^{(NS)} &=& - c_1 {\psi^*}^2-\frac{g c'_7}{\left(g+\phi^*\right)^2}, \nonumber \\
\mathcal{M^*}_{12}^{(NS)} &=& - 2 c_1 \phi^*\psi^*, \nonumber \\
\mathcal{M^*}_{13}^{(NS)} &=& 2 c_3 \eta^*, \nonumber \\
\mathcal{M^*}_{21}^{(NS)} &=& c_2 {\psi^*}^2, \nonumber \\
\mathcal{M^*}_{22}^{(NS)} &=& 2 c_2 \phi^*\psi^* - c_4, \\
\mathcal{M^*}_{23}^{(NS)} &=& \mathcal{M}_{32}^{(NS)} = 0, \nonumber \\
\mathcal{M^*}_{31}^{(NS)} &=& c_5, \nonumber \\
\mathcal{M^*}_{33}^{(NS)} &=& -c_6, \nonumber \\
\mathcal{M^*}_{rs}^{(SP)} &=& d_r \delta_{rs} \nonumber,
\EE
and those of matrix $\mathcal B^*$ are
\BE \label{eq:Bfp}
\mathcal{B^*}_{11}^{(NS)} &=& c_1 \phi^*{\psi^*}^2 + c_3 \eta^{*2} + c'_7 \frac{\phi^*}{g+\phi^*}, \nonumber \\
\mathcal{B^*}_{22}^{(NS)} &=& c_8 + c_2 \phi^*{\psi^*}^2 + c_4 \psi^*, \nonumber \\
\mathcal{B^*}_{33}^{(NS)} &=& c_5 \phi^* + c_6 \eta^*, \nonumber \\
\mathcal{B^*}_{rs}^{(NS)} &=& 0, \quad \text{with } r \ne s, \\
\mathcal{B^*}_{11}^{(SP)}&=& - 2d_1 \phi^*, \nonumber \\
\mathcal{B^*}_{22}^{(SP)}&=& - 2d_2 \psi^*, \nonumber\\
\mathcal{B^*}_{33}^{(SP)}&=& -2d_3 \eta^*, \nonumber \\
\mathcal{B^*}_{rs}^{(SP)}&=& 0, \quad \text{with } r \ne s. \nonumber
\EE
 
\section{Analysis of the Power Spectra} \label{sec:B}

The Fokker-Planck equation \eqref{eq:fp2}, with the the matrices evaluated at the fixed point, describes fluctuations about the fixed point, and is equivalent to the Langevin equation~\cite{vanKampen}:
\BE \label{eq:lang}
\frac{d \xi_{r,i}}{d\tau} &=& \sum_{s=1}^3 \sum_{j=1}^\Omega \mathcal{M}^*_{rs,ij} \xi_{s,j} + \chi_{r, i} \\
&=& \sum_{s=1}^3 \sum_{j=1}^\Omega \left( \mathcal{M^*}_{rs}^{(NS)}\delta_{ij} + \mathcal{M^*}_{rs}^{(SP)} \Delta_{ij}\right) \xi_{s,j} + \chi_{r, i} \nonumber.
\EE
The Gaussian white noises $\chi_{r, i}$ have zero mean and correlator:
\be \label{eq:corr}
\langle \chi_{r, i}(\tau) \chi_{s, j}(\tau') \rangle = \mathcal B_{rs,ij} \delta(\tau - \tau').
\ee
Equation \eqref{eq:lang} generalises \eqref{eq:linstab} to include stochastic fluctuations. In solving Eq.~\eqref{eq:lang}, we again make use of the transforms \eqref{eq:trans}. We express the $\xi_{r,i}$ and the associated noise in terms of their transformed analogs. Collecting each term, except the noise, to the left-hand side of the equation yields:
\be \label{eq:slinstabtrans}
\left(- \mathrm i \omega \mathcal I - \mathcal {M^*}^{(NS)} - \mathcal {M^*}^{(SP)} \Lambda^{(\alpha)}  \right) \cdot \begin{pmatrix} \tilde{\xi}_{1,\alpha} \\ \tilde{\xi}_{2,\alpha} \\  \tilde{\xi}_{3,\alpha} \end{pmatrix} = \begin{pmatrix} \tilde{\chi}_{1,\alpha} \\ \tilde{\chi}_{2,\alpha} \\  \tilde{\chi}_{3,\alpha} \end{pmatrix},
\ee 
where $\mathcal I$ is the $3\times 3$ identity matrix. By introducing $\mathcal F^{(\alpha)} = -\mathrm i \omega \mathcal I - \mathcal {M^*}^{(NS)} - \mathcal {M^*}^{(SP)} \Lambda^{(\alpha)}$ the solution of Eq.~\eqref{eq:slinstabtrans} may be written as:
\be \label{eq:sol}
\tilde{\xi}_{r,\alpha} = \sum_{s=1}^3 \mathcal F^{-1}_{rs} \tilde{\chi}_{s,\alpha},
\ee
where we have omitted the $\alpha$ index on $\mathcal F$ for clarity. We now insert Eq.~\eqref{eq:sol} into Eq.~\eqref{eq:ps} to obtain an expression for the power spectra:
\be \label{eq:analps0}
	P_r(\omega, \Lambda^{(\alpha)}) = \sum_{s,l = 1}^3 \mathcal F^{-1}_{rl} \langle \tilde \chi_{l,\alpha} \tilde \chi^c_{s,\alpha} \rangle \mathcal F^{-1\dagger}_{sr}.
\ee
The symbol $\dagger$ signifies the adjoint operator, here equivalent to the conjugate transpose operator. We now need to express $\langle \tilde \chi_{l,\alpha} \tilde \chi^c_{s,\alpha} \rangle$ in terms of known quantities. We begin by transforming Eq.~\eqref{eq:corr} using the inverse transform \eqref{eq:trans}, which leads to
\be \label{eq:elem}
 \langle \tilde \chi_{l,\alpha} \tilde \chi^c_{s,\alpha} \rangle = 2 \pi \sum_{i,j=1}^\Omega v_i^{(\alpha)} v_j^{(\alpha)} \mathcal B^*_{ls, ij}.
\ee
The dependence on the Laplacian eigenvectors can be eliminated using the fact that they are orthonormal and complete:
\BE \label{eq:lapprop}
\sum_{i=1}^{\Omega} v_i^{(\alpha)} v_i^{(\alpha')} = \delta_{\alpha\alpha'},\quad
\sum_{\alpha=1}^{\Omega} v_i^{(\alpha)} v_j^{(\alpha)} = \delta_{ij}.
\EE	
To do so, we substitute the decomposition \eqref{eq:decom} into Eq.~\eqref{eq:elem}, then use the above properties to arrive at:
\be \label{eq:elem2}
 \langle \tilde \chi_{l,\alpha} \tilde \chi^c_{s,\alpha} \rangle = 2 \pi \left( \mathcal {B^*}^{(NS)}_{ls} + \mathcal {B^*}^{(SP)}_{ls} \Lambda^{(\alpha)} \right).
\ee
The right-hand side of Eq.~(\ref{eq:elem2}) is known through expressions \eqref{eq:Bfp}, and so $\langle \tilde \chi_{l,\alpha} \tilde \chi^c_{s,\alpha} \rangle$ can be found. By substituting  Eq.~\eqref{eq:elem} into Eq.~\eqref{eq:analps0} we arrive at the final formula for the power spectra:
\BE \label{eq:analps}
	P_r(\omega, \Lambda^{(\alpha)}) &=& \sum_{s,l = 1}^3 \mathcal F^{-1}_{rl} \left( \mathcal {B^*}^{(NS)}_{ls} + \mathcal {B^*}^{(SP)}_{ls} \Lambda^{(\alpha)} \right) \mathcal F^{-1\dagger}_{sr} \nonumber\\
	&=& \left(\mathcal F^{-1} \left( \mathcal {B^*}^{(NS)} + \mathcal {B^*}^{(SP)} \Lambda^{(\alpha)} \right) \mathcal F^{-1\dagger}\right)_{rr}. \nonumber \\
\EE

\end{document}